# New Sequence Sets with Zero-Correlation Zone

Xiangyong Zeng, Lei Hu, Qingchong Liu, *Member*

**Abstract**

A method for constructing sets of sequences with zero-correlation zone (ZCZ sequences) and sequence sets with low cross correlation is proposed. The method is to use families of short sequences and complete orthogonal sequence sets to derive families of long sequences with desired correlation properties. It is a unification of works of Matsufuji and Torii *et al.*, and there are more choices of parameters of sets for our method. In particular, ZCZ sequence sets generated by the method can achieve a related ZCZ bound. Furthermore, the proposed method can be utilized to derive new ZCZ sets with both longer ZCZ and larger set size from known ZCZ sets. These sequence sets are applicable in broadband satellite IP networks.

L. Hu's work is supported in part by the National Science Foundation of China (NSFC) under Grants No.60373041 and No.90104034. Q. Liu's work is supported in part by NSF under grant CNS-0435341.

X. Zeng is with the Faculty of Mathematics and Computer Science, Hubei University, 11 Xueyuan Road, Wuhan 430062, P. R. China, email: xzeng@hubu.edu.cn.

L. Hu is with the Graduate School of Chinese Academy of Sciences, 19A Yuquan Road, Beijing 100049, P. R. China.

Q. Liu is with Department of Electrical and System Engineering, Oakland University, Rochester, MI 48309, USA.

# I. INTRODUCTION

Recent practice in broadband satellite IP networks has demanded sequences of low correlation in a small detection aperture [1]-[3]. In a broadband satellite IP network, there is always a control channel broadcasting the system time along with other system information [1], [2]. A user terminal listens to the control channel, adjusts its timing accordingly so that its packet arrives at the central receiver within a few symbols [1]. Because a broadband satellite IP network has to support a very large number of terminals, which can be over 10 million, there can be many simultaneous transmissions from different terminals [1], [2]. Sequences need to be designed as signature sequences to differentiate terminals, which are often called unique words [4]. Such sequences need to have low values of both autocorrelation and crosscorrelation within the detection aperture. These requirements translate to the design of sequences of zero-correlation zone (ZCZ), or sequences of low correlation in an aperture.

Polyphase ZCZ sequence sets were constructed by Suehiro for the first time [5]. The concept of zero-correlation zone (ZCZ) was introduced in [6]. Several classes of ZCZ sequences were derived based on complementary pairs [5], [7], [8]. The construction was extended in [9], [10] by employing complementary sets. Binary sequence sets make hardware implementation much easier than polyphase sequences. However, binary ZCZ sequence sets can not achieve the upper bound of set size [11]. Sets of ternary ZCZ sequences with entries in $\{\pm 1, 0\}$ can achieve the upper bound. Their hardware implementation is similar to binary sequences. Several classes of ternary ZCZ sequences have been constructed [12]-[16]. Important progress was also achieved in the design of sequences using short sequences to construct a set of long sequences with the desired correlation properties [17]-[20]. By extending the concept of ZCZ to the two-dimensional case, families of ZCZ arrays where the one-dimensional ZCZ becomes a rectangular ZCZ can also be synthesized [21]-[23]. More references can be found in [24]. In addition, some polyphase sequences with low correlation zone have been constructed in [25]-[27].

In this correspondence, we present a general method for constructing sets of ZCZ sequences and sequence sets with low cross correlation. The original idea of this method was proposed by Gong in 1995 [17], where she obtained new low-correlation sequence sets by interleaving two ideal sequences with two-level periodic autocorrelation property [18]. Different from Gong's approach, we use two sets of sequences to replace Gong's ideal sequences and their shift equivalent sequences. One set of sequences is complete orthogonal, while the other set can consist of either shift equivalences of a fixed perfect sequence or sequences inequivalent and with some good correlation properties.

Our method also unifies the constructions presented in [19], [20]. The ZCZ sequence sets constructed in [19], [20] can be constructed by our method. Furthermore, there are more choices of parameters of sequence sets for our method. In particular, ZCZ sequence sets generated by our method can achieve the bound of set size, while those constructed in [20] can not. Furthermore, our method can be utilized to derive new ZCZ sets with both longer ZCZ and larger set size from known ZCZ sets.

The remainder of this correspondence is organized as follows. Section II gives some definitions and introduce a basic method for constructing sequences. Sets of sequences with desired correlation properties are constructed by using the proposed method in Section III and IV. Section V concludes the study.

## II. Preliminaries and A Basic Construction Method

Throughout the correspondence all sequences, except shift sequences, have entries in the complex field.

### A. Some Definitions

Let $\mathcal{S} = \{\mathbf{s}_0, \mathbf{s}_1, \cdots, \mathbf{s}_{M-1}\}$ be a set of $M$ sequences of length $N$, where

$$\mathbf{s}_h = (s_{h,0}, s_{h,1}, \cdots, s_{h,N-1}) \text{ for } 0 \leq h < M. \tag{1}$$

The *periodic cross-correlation function* $R_{\mathbf{s}_h,\mathbf{s}_k}(\tau)$ of $\mathbf{s}_h$ and $\mathbf{s}_k$, $0 \leq h, k \leq M-1$, is defined as

$$R_{\mathbf{s}_h,\mathbf{s}_k}(\tau) = \sum_{i=0}^{N-1} s_{h,i} \cdot s^*_{k,i+\tau}, \quad \tau = 0, 1, \cdots \tag{2}$$

where the symbol $*$ denotes a complex conjugate and the subscript addition is performed modulo $N$. If $h = k$, then $R_{\mathbf{s}_h,\mathbf{s}_k}(\tau)$ is called the *periodic autocorrelation function* of $\mathbf{s}_h$, denoted by $R_{\mathbf{s}_h}(\tau)$.

The *energy* of $\mathbf{s}_h$ is defined as

$$E_{\mathbf{s}_h} = \sum_{i=0}^{N-1} |s_{h,i}|^2. \tag{3}$$

$\mathbf{s}_h$ is called a *perfect sequence* if

$$R_{\mathbf{s}_h}(\tau) = \begin{cases} E_{\mathbf{s}_h}, & \text{if } \tau \equiv 0 \bmod N \\ 0, & \text{otherwise} \end{cases} \tag{4}$$

The sequence $\mathbf{s}_h$ is called a $p$*-phase sequence* if for any $0 \le i \le N-1$, $\mathbf{s}_{h,i} = \exp(\frac{2l_i\pi j}{p})$ for some $0 \le l_i < p$, where $j^2 = -1$. In the case of $p = 2$, the sequence is called a *binary sequence*.

The sequence set $\mathcal{S}$ is said to be *orthogonal* if the set has the following characteristic:

$$R_{\mathbf{s}_h,\mathbf{s}_k}(\tau) = \begin{cases} E_{\mathbf{s}_h}, & \text{if } h = k, \tau = 0 \\ 0, & \text{if } h \ne k, \tau = 0 \end{cases} \tag{5}$$

for any $0 \le h, k < M$. In the case of $N = M$, the set $\mathcal{S}$ is said to be *complete orthogonal*.

$\mathcal{S}$ is said to be an $(N, M; Zcz)$*-ZCZ set* and has a zero-correlation zone of size $Zcz$ if

$$R_{\mathbf{s}_h,\mathbf{s}_k}(\tau) = \begin{cases} E_{\mathbf{s}_h}, & \text{if } h = k, \tau = 0 \\ 0, & \text{if } h \ne k, \tau = 0 \\ 0, & \text{if } 1 \le |\tau| \le Zcz \end{cases} \tag{6}$$

for any $0 \le h, k < M$. Thus, for any $\mathbf{s}_h$, $\mathbf{s}_k$ in $\mathcal{S}$,

$$R_{\mathbf{s}_h}(\tau) = 0 \text{ for } 1 \le |\tau| \le Zcz \quad \text{and} \quad R_{\mathbf{s}_h,\mathbf{s}_k}(\tau) = 0 \text{ for } 0 \le |\tau| \le Zcz \text{ and } h \ne k.$$

Such a set is also said to be $Z$*-orthogonal* [10].

A mathematical bound

$$Zcz \le \frac{N}{M} - 1 \tag{7}$$

holds for an arbitrary $(N, M; Zcz)$-ZCZ set [11]. It is derived from the Welch lower bound [28], and is later introduced by Matsufuji *et al.* with a more simple proof [19].

Set

$$\delta = \max|R_{\mathbf{s}_h,\mathbf{s}_k}(\tau)|, \tag{8}$$

where $0 \le| \tau |< N$, $0 \le h, k < M$ and the maximization excludes the case of $\tau = 0$ and $h = k$. We call $\delta$ the *maximal correlation* of $\mathcal{S}$ and $\mathcal{S}$ an $[N, M; \delta]$ sequence set.

For the sequence $\mathbf{s}_h = (s_{h,0}, s_{h,1}, \cdots, s_{h,N-1})$, the *left shift operator* $L$ on $\mathbf{s}_h$ is defined as

$$L(\mathbf{s}_h) = (s_{h,1}, \cdots, s_{h,N-1}, s_{h,0}). \tag{9}$$

For any $i > 0$, iteratively define

$$L^i(\mathbf{s}_h) = L(L^{i-1}(\mathbf{s}_h)) \tag{10}$$

where $L^0(\mathbf{s}_h) = \mathbf{s}_h$. Sequences $\mathbf{s}_h$ and $\mathbf{s}_k$ are called *(cyclically) shift equivalent* if there exists an integer $k$ such that $\mathbf{s}_h = L^k(\mathbf{s}_k)$.

For a sequence set $\mathcal{A} = \{\mathbf{a}_0, \mathbf{a}_1, \cdots, \mathbf{a}_{n-1}\}$, define

$$L^i(\mathcal{A}) = \{L^i(\mathbf{a}_0), L^i(\mathbf{a}_1), \cdots, L^i(\mathbf{a}_{n-1})\}. \tag{11}$$

Two sequence sets $\mathcal{A}$ and $\mathcal{B}$ are called shift equivalent if there exists an integer $i$ such that $\mathcal{A} = L^i(\mathcal{B})$. When all sequences in the set $\mathcal{A}$ are shift equivalent to a fixed sequence $\mathbf{a}$, namely,

$$\mathcal{A} = \{L^{e_0}(\mathbf{a}), L^{e_1}(\mathbf{a}), \cdots, L^{e_{n-1}}(\mathbf{a})\}, \tag{12}$$

the integer sequence $\mathbf{e} = (e_0, e_1, \cdots, e_{n-1})$ is called a *shift sequence*.

The following notations are used in the rest of this correspondence:

- $m|n$: the integer $n$ is a multiple of $m$;
- $\gcd(p, q)$: the greatest common divisor of integers $p$ and $q$.
- $\lfloor z \rfloor$: the largest integer not exceeding $z$.

### *B. A Basic Construction of Sequence Set*

A basic method to construct a sequence set is given in the following procedure.

Procedure 1:

(1) Let $\mathcal{A}$ =$\{\mathbf{a}_0, \mathbf{a}_1, \cdots, \mathbf{a}_{n-1}\}$ be an ordered set of sequences, where $\mathbf{a}_i = (a_{i,0}, a_{i,1}, \cdots, a_{i,m-1})$. Choose a complete orthogonal sequence set $\mathcal{B} = \{\mathbf{b}_0, \mathbf{b}_1, \cdots, \mathbf{b}_{n-1}\}$, where $\mathbf{b}_i = (b_{i,0}, b_{i,1}, \cdots, b_{i,n-1})$.

(2) Let $U$ be the $m \times n$ matrix whose $j$-th column sequence is $\mathbf{a}_j$. Listing all entries of $U$ row by row (from left to right and from top to bottom), we obtain a sequence of length $mn$,

$$\mathbf{u} = (u_0, u_1, \cdots, u_{mn-1}), \tag{13}$$

which is called the sequence associated with the ordered set $\mathcal{A}$, and $U$ is called its matrix form.

(3) Set $\mathbf{s}_h = (s_{h,0}, s_{h,1}, \cdots, s_{h,mn-1})$, $0 \leq h < n$, where

$$s_{h,i} = u_i b_{h,\mathrm{imod}n} \tag{14}$$

for $0 \leq i < mn$.

(4) The sequence set $\mathcal{S} = \mathcal{S}(\mathcal{A}, \mathcal{B})$ is defined as $\mathcal{S} = \{\mathbf{s}_0, \mathbf{s}_1, \cdots, \mathbf{s}_{n-1}\}$.

*Example 1:* Let $\mathcal{A} = \{\mathbf{a}_0, \mathbf{a}_1\}$, where $\mathbf{a}_0 = (1, -1, -1)$, $\mathbf{a}_1 = (1, 1, -1)$, and $\mathcal{B}$ consist of two row sequences of the Hadamard matrix

$$H = \begin{pmatrix} 1 & 1 \\ 1 & -1 \end{pmatrix}. \tag{15}$$

By Procedure 1, $\mathcal{S} = \{\mathbf{s}_0, \mathbf{s}_1\}$ where

$$\mathbf{s}_0 = (1, 1, -1, 1, -1, -1), \ \mathbf{s}_1 = (1, -1, -1, -1, -1, 1).$$

The original idea of this construction was proposed by Gong in 1995 [17], and later she employed it to construct more sequence sets [18]. In her construction, she assumes that Eq. (12) and Eq. (16) below

$$\mathcal{B} = \{\mathbf{b}, L^1(\mathbf{b}), \cdots, L^{n-1}(\mathbf{b})\} \tag{16}$$

hold for two ideal two-level autocorrelation sequences $\mathbf{a}$ and $\mathbf{b}$ and a shift sequence $\mathbf{e} = (e_0, e_1, \cdots, e_{n-1})$.

A similar construction was given by Torii *et al.* in [20], where the divisibility condition $n|m$ or $m|n$ was required. However, none of the sequence sets generated by that construction achieves the bound (7). Recently, a new method for generating sets of polyphase ZCZ sequences achieving the bound (7) was given in [19].

Procedure 1 unifies the constructions in [19], [20]. In Sections III and IV, we concentrate on constructing sequence sets with good correlation property.

The following formula on correlation of sequences constructed from Procedure 1 is a basis on which we prove the results in the correspondence.

Let $0 \leq \tau < mn$ and write $\tau = rn + s$ with $0 \leq r < m$ and $0 \leq s < n$.

*Proposition 1:* For $0 \leq h, k < n$,

$$R_{\mathbf{s}_h, \mathbf{s}_k}(\tau) = \sum_{j=0}^{n-1} d_i R_{\mathbf{a}_i, \mathbf{a}_{s+i-\varphi(s+i)n}}(r + \varphi(s+i)) \tag{17}$$

where $d_i = b_{h,i} b^*_{k,s+i-\varphi(s+i)n}$ and $\varphi(s+i)$ is 0 if $s+i<n$ and is 1 otherwise.

The proof of this proposition is presented in Appendix I.

## III. A Class of ZCZ Sequence Sets

In this section, we first use the procedure in Section II-B to construct new ZCZ sequences. Next, we give a brief comment on some known constructions of perfect sequences and complete orthogonal sequence sets. Finally, we study how to construct sets of long ZCZ sequences with both longer ZCZ and larger set size, based on known sets of short ZCZ sequences.

### *A. Construction of ZCZ Sequences Based on Perfect Sequences*

In this subsection all sequences of the set $\mathcal{A}$ are assumed to be shift equivalent, i.e., Eq. (12) holds. In this case, Eq. (17) is further simplified to

$$\begin{aligned} R_{\mathbf{s}_h,\mathbf{s}_k}(\tau) &= \sum_{i=0}^{n-1} d_i R_{\mathbf{a}_i,\mathbf{a}_{s+i-\varphi(s+i)n}}(r+\varphi(s+i)) \\ &= \sum_{i=0}^{n-1} d_i R_{\mathbf{a}}(e_{i+s}-e_i+r) \end{aligned} \tag{18}$$

where we define $e_i = 1 + e_{i-n}$ for subscript $i \geq n$.

By Eq. (18), if Eq. (16) holds, and if both $\mathbf{a}$ and $\mathbf{b}$ have an ideal two-level autocorrelation function, then the correlation value $R_{\mathbf{s}_h,\mathbf{s}_k}$ can be completely determined by the shift sequence $\mathbf{e} = (e_0, e_1, \cdots, e_{n-1})$. Based on this nice observation, Gong [18] presented two methods to construct the shift sequences with desirable combinational property, which guarantees the constructed signal sets have low correlation property.

Gong's idea is extended to construct families of ZCZ sequences in this subsection. If $\mathbf{a}$ is a perfect sequence and $\mathcal{B}$ is complete orthogonal, then we can choose the shift sequences such that the sequences sets as described in Theorems 1 and 2 have desirable ZCZ property.

Set

$$r_0 = \min\{m + e_0 - e_{n-1}, e_{i+1} - e_i - 1 \mid 0 \leq i \leq n-2\}. \tag{19}$$

*Theorem 1:* Let $m \geq n$ and

$$0 \leq e_0 < e_1 < \cdots < e_{n-1} < m.$$

If $\mathbf{a} = (a_0, a_1, \cdots, a_{m-1})$ is a perfect sequence, then $\mathcal{S}$ is an $(mn, n; r_0 n + n - 2)$-ZCZ set. Furthermore,

(1) When $n|(m+1)$ and $e_i = \frac{i(m+1)}{n}$ for $0 \leq i < n$, $\mathcal{S}$ is an $(mn, n; m-1)$-ZCZ set.

(2) When $n|m$ and

$$\frac{(i_1 - i_2)m}{n} \leq e_{i_1} - e_{i_2} \leq \frac{(i_1 - i_2)m}{n} + 1 \tag{20}$$

for any $0 \leq i_2 < i_1 \leq n-1$, $\mathcal{S}$ is an $(mn, n; m-2)$-ZCZ set. Each of the shift sequences $\mathbf{e}^{(0)}, \cdots, \mathbf{e}^{(n-2)}$, and $\mathbf{e}^{(n-1)}$, defined by

$$\mathbf{e}^{(i)} = (0, \frac{m}{n}, \cdots, \frac{(n-i-1)m}{n}, \frac{(n-i)m}{n} + 1, \frac{(n-i+1)m}{n} + 1, \cdots, \frac{(n-1)m}{n} + 1), \tag{21}$$

satisfies Eq. (20).

The proof of this theorem is presented in Appendix II.

When $n|(m+1)$, we obtain ZCZ sequence sets which achieve the bound (7) by Theorem 1(1). The polyphase ZCZ sets constructed in [6] also attain this bound, however, the number of phases is equal to the length of the sequences in the set. Applying Theorem 1(1), we can obtain polyphase ZCZ sets with fewer number of phases than the sequence length. Precisely, if $\mathbf{a}$ and those sequences in $\mathcal{B}$ are of $p$- and $q$-phase respectively, then the sequences in $\mathcal{S}$ are of $\frac{pq}{\gcd(p,q)}$-phase. This number of phases may be independent on the sequence length. A disadvantage is that the condition $n|m+1$ restricts us in the construction of binary, three-phase, or quadriphase ZCZ sequence sets according to Theorem 1(1).

When $n|m$, the ZCZ sets constructed by Theorem 1(2) do not achieve the bound (7).

*Example 2:* Let $m = 9$, $n = 2$, and $\mathbf{e} = (0, 5)$. Suppose that

$$\mathbf{a} = (0, 0, 0, 0, 1, 2, 0, 2, 1)$$

is a three-phase perfect sequence where $0, 1$ and $2$ represent 1, $\exp(\frac{2\pi j}{3})$ and $\exp(\frac{4\pi j}{3})$, respectively, and $\mathcal{B}$ is the same as in Example 1. A six-phase $(18, 2; 8)$-ZCZ set $\mathcal{S}$ shown as

$$\mathcal{S} = \{(040004022040004022), (010301052343034325)\}$$

is obtained by applying Theorem 1(1), where an entry $i$ represents a six-phase complex number $\exp(\frac{2i\pi j}{6})$. The absolute values of autocorrelation functions $R_{\mathbf{s}_0}(\tau)$ and $R_{\mathbf{s}_1}(\tau)$ are given by

$$|R_{\mathbf{s}_0}(\tau)| = |R_{\mathbf{s}_1}(\tau)| = (18, 0, 0, 0, 0, 0, 0, 0, 0, 18, 0, 0, 0, 0, 0, 0, 0, 0).$$

The absolute values of cross-correlation functions $R_{\mathbf{s}_0,\mathbf{s}_1}(\tau)$ and $R_{\mathbf{s}_1,\mathbf{s}_0}(\tau)$ are given by

$$|R_{\mathbf{s}_0,\mathbf{s}_1}(\tau)| = |R_{\mathbf{s}_1,\mathbf{s}_0}(\tau)| = (0, 0, 0, 0, 0, 0, 0, 0, 0, 0, 0, 0, 0, 0, 0, 0, 0, 0).$$

*Example 3:* Consider quadriphase sequence sets. Let $0, 1, 2$ and $3$ represent $+1, +j, -1$ and $-j$, respectively. In the above terminology, the $(64, 4; 14)$-ZCZ set $\mathcal{C}_1$ in P. 561 of [20] is constructed from the perfect quadriphase sequence $\mathbf{a} = (0000012302020321)$, a complete orthogonal sequence set $\mathcal{B} = \{0000, 0123, 0202, 0321\}$ and the shift sequence $\mathbf{e}^{(0)} = (0, 4, 8, 12)$. By applying Theorem 1(2), there exist other 3 quadriphase $(64, 4; 14)$-ZCZ sets which are constructed from $\mathbf{e}^{(1)}, \mathbf{e}^{(2)}$ and $\mathbf{e}^{(3)}$, respectively. They are not shift equivalent to $C_1$ and the four sets are listed as follows:

(1) $\mathbf{e}^{(0)} = (0, 4, 8, 12)$

$\mathbf{s}_0 = (0000012302020321000012302020321000002301020221030000301220201032)$

$\mathbf{s}_1 = (0123020203210000012313132103333012320200321222201233131210311111)$

$\mathbf{s}_2 = (0202032100000123020210322222301202022103000023010202321022221230)$

$\mathbf{s}_3 = (0321000001230202032111112301313103212222012320200321333323011313)$

(2) $\mathbf{e}^{(1)} = (0, 4, 8, 13)$

$\mathbf{s}_0 = (0003012202010320000012302020321000012302020321000002301020221030)$

$\mathbf{s}_1 = (0122020103200003012313132103333012020210322222301213133210111113)$

$\mathbf{s}_2 = (0201032000030122020210322222301202032100000123020200321222201232)$

$\mathbf{s}_3 = (0320000301220201032111112301313103222223012020210323333123031311)$

(3) $\mathbf{e}^{(2)} = (0, 4, 9, 13)$

$\mathbf{s}_0 = (0023010202210300003012202010320000012302020321000012302020321000)$

$\mathbf{s}_1 = (0102022103000023011313032133332301202021032222230131310321111123)$

$\mathbf{s}_2 = (0221030000230102023210222212300202032100000123020210322222301202)$

$\mathbf{s}_3 = (0300002301020221031111012331312103222223012020210333330123131321)$

(4) $\mathbf{e}^{(3)} = (0, 5, 9, 13)$

$\mathbf{s}_0 = (0123020203210000023010202210300003012202010320000012302020321000)$

$\mathbf{s}_1 = (0202032100000123031311032333312300202321022221230131310321111123)$

$\mathbf{s}_2 = (0321000001230202003212222012320201032000030122020210322222301202)$

$\mathbf{s}_3 = (0000012302020321011113012131332102222123002023210333330123131321)$

For each sequence set, the absolute values of autocorrelation functions $R_{\mathbf{s}_0}(\tau)$, $R_{\mathbf{s}_1}(\tau)$, $R_{\mathbf{s}_2}(\tau)$ and $R_{\mathbf{s}_3}(\tau)$ are given by

$$
\begin{aligned}
|R_{\mathbf{s}_0}(\tau)| = |R_{\mathbf{s}_1}(\tau)| = |R_{\mathbf{s}_2}(\tau)| = |R_{\mathbf{s}_3}(\tau)| = (&64, 0, 0, 0, 0, 0, 0, 0, 0, \\
&0, 0, 0, 0, 0, 0, 48, 0, 0, 0, 16, 0, 0, 0, 0, 0, 0, 0, 0, 0, 0, 32, 0, 0, 0, 32, 0, \\
&0, 0, 0, 0, 0, 0, 0, 0, 0, 16, 0, 0, 0, 48, 0, 0, 0, 0, 0, 0, 0, 0, 0, 0, 0, 0, 0, 0)
\end{aligned}
$$

and the absolute values of cross-correlation functions are given by

$$
\begin{aligned}
&|R_{\mathbf{s}_0,\mathbf{s}_1}(\tau)| = |R_{\mathbf{s}_1,\mathbf{s}_0}(\tau)| = |R_{\mathbf{s}_0,\mathbf{s}_3}(\tau)| = |R_{\mathbf{s}_3,\mathbf{s}_0}(\tau)| = |R_{\mathbf{s}_1,\mathbf{s}_2}(\tau)| = \\
&|R_{\mathbf{s}_2,\mathbf{s}_1}(\tau)| = |R_{\mathbf{s}_2,\mathbf{s}_3}(\tau)| = |R_{\mathbf{s}_3,\mathbf{s}_2}(\tau)| = (0, 0, 0, 0, 0, 0, 0, 0, 0, 0, 0, 0, \\
&0, 0, 0, 16, 0, 0, 0, 16, 0, 0, 0, 0, 0, 0, 0, 0, 0, 0, 22.62, 0, 0, 0, 22.62, 0, \\
&0, 0, 0, 0, 0, 0, 0, 0, 0, 16, 0, 0, 0, 16, 0, 0, 0, 0, 0, 0, 0, 0, 0, 0, 0, 0, 0, 0)
\end{aligned}
$$

and

$$
\begin{aligned}
|R_{\mathbf{s}_0,\mathbf{s}_2}(\tau)| = |R_{\mathbf{s}_2,\mathbf{s}_0}(\tau)| = |R_{\mathbf{s}_1,\mathbf{s}_3}(\tau)| = |R_{\mathbf{s}_3,\mathbf{s}_1}(\tau)| = (&0, 0, 0, 0, 0, 0, 0, 0, \\
&0, 0, 0, 0, 0, 0, 0, 16, 0, 0, 0, 16, 0, 0, 0, 0, 0, 0, 0, 0, 0, 0, 0, 0, 0, 0, 0, 0, \\
&0, 0, 0, 0, 0, 0, 0, 0, 0, 16, 0, 0, 0, 16, 0, 0, 0, 0, 0, 0, 0, 0, 0, 0, 0, 0, 0, 0).
\end{aligned}
$$

*Theorem 2:* Let $m \geq n$, $\mathbf{a} = (a_0, a_1, \cdots, a_{m-1})$ be a perfect sequence, and $\mathbf{e} = (e_0, e_1, \cdots, e_{n-1})$ be a shift sequence defined by $e_i = m - 1 - i$. Then $\mathcal{S}$ is an $(mn, n; n \bmod m)$-ZCZ set. Furthermore, when $m = n + 1$, $\mathcal{S}$ is an $(mn, n; m - 1)$-ZCZ set.

The proof of Theorem 2 is presented in Appendix III.

For $m$ and $n$, by choosing appropriate shift sequences, we can construct more sequence sets with parameters different from those constructed in [20].

*Example 4:* Suppose $m = 16$, $n = 12$,

$$\mathbf{e} = (15, 14, 13, 12, 11, 10, 9, 8, 7, 6, 5, 4),$$

and $\mathbf{a}$ is the perfect sequence shown in Example 3. The complete orthogonal sequence set $\mathcal{B}$ is supposed to consist of row sequences of the following matrix:

$$\begin{pmatrix}
+ & + & - & + & + & + & - & - & - & + & - & - \\
+ & - & + & + & + & - & - & - & + & - & - & + \\
+ & + & + & + & - & - & - & + & - & - & + & - \\
+ & + & + & - & - & - & + & - & - & + & - & + \\
+ & + & - & - & - & + & - & - & + & - & + & + \\
+ & - & - & - & + & - & - & + & - & + & + & + \\
+ & - & - & + & - & - & + & - & + & + & + & - \\
+ & - & + & - & - & + & - & + & + & + & - & - \\
+ & + & - & - & + & - & + & + & + & - & - & - \\
+ & - & - & + & - & + & + & + & - & - & - & + \\
+ & - & + & - & + & + & + & - & - & - & + & - \\
+ & + & + & + & + & + & + & + & + & + & + & +
\end{pmatrix}$$

where the symbols $+$ and $-$ represent $+1$ and $-1$, respectively. By Procedure 1, we obtain a quadriphase $(192, 12; 12)$-ZCZ set. Since the length of these sequences is very long and the set size is large, we only list the first two sequences in $\mathcal{S}$ as follows:

$$\begin{aligned}
\mathbf{s}_0 = (&121020021232010302202303003230020010002123202221002012120002 0\\
&020010122201020003010022120002303203230002232120301002221012012102\\
&2203002232122202320003232202202220303202230002010302223220221012 2)\\
\mathbf{s}_1 = (&103022023030032300200101021232022212020121200023020010122200 0\\
&200030100221200023032002300022321223010022210100121022203032232122\\
&2023200032322022122203032022000020103022032202210122021020021232 0)
\end{aligned}$$

The absolute values of autocorrelation function of $\mathrm{s}_0$ are given by

$$\begin{aligned}
|R_{\mathbf{s}_0}(\tau)| = (&192, 0, 0, 0, 0, 0, 0, 0, 0, 0, 0, 0, 0, 16, 0, 0, 0, 0, 0, 0, 0, 0, 0, 0, 0, 0, 32, 0,\\
&0, 0, 0, 0, 0, 0, 0, 0, 0, 0, 0, 16, 0, 0, 0, 0, 0, 0, 0, 0, 0, 16, 0, 0, 64, 0, 0, 0, 0, 0, 0, 0, 0,\\
&0, 32, 0, 0, 16, 0, 0, 0, 0, 0, 0, 0, 0, 0, 16, 0, 0, 32, 0, 0, 0, 0, 0, 0, 0, 0, 0, 0, 0, 0, 80, 0,\\
&0, 0, 0, 0, 0, 0, 0, 0, 80, 0, 0, 0, 0, 0, 0, 0, 0, 0, 0, 0, 0, 32, 0, 0, 16, 0, 0, 0, 0, 0, 0, 0, 0,\\
&0, 16, 0, 0, 32, 0, 0, 0, 0, 0, 0, 0, 0, 0, 64, 0, 0, 16, 0, 0, 0, 0, 0, 0, 0, 0, 0, 16, 0, 0, 0, 0,\\
&0, 0, 0, 0, 0, 0, 0, 0, 32, 0, 0, 0, 0, 0, 0, 0, 0, 0, 0, 0, 0, 16, 0, 0, 0, 0, 0, 0, 0, 0, 0, 0, 0, 0).
\end{aligned}$$

The other sequences in $\mathcal{S}$ have the same autocorrelation property concerning the ZCZ. The absolute values of cross-correlation function of $\mathbf{s}_0$ and $\mathbf{s}_1$ are given by

$$\begin{aligned}|R_{\mathbf{s}_0,\mathbf{s}_1}(\tau)| = (&0,0,0,0,0,0,0,0,0,0,0,0,0,48,0,0,0,0,0,0,0,0,0,0,0,0,0,0,0,\\ &0,0,0,0,0,0,0,0,0,0,48,0,0,0,0,0,0,0,0,0,16,0,0,0,0,0,0,0,0,0,0,0,0,\\ &0,0,0,48,0,0,0,0,0,0,0,0,0,16,0,0,64,0,0,0,0,0,0,0,0,0,64,0,0,16,0,0,\\ &0,0,0,0,0,0,0,16,0,0,32,0,0,0,0,0,0,0,0,0,0,0,0,48,0,0,0,0,0,0,0,0,0,\\ &48,0,0,0,0,0,0,0,0,0,0,0,0,32,0,0,16,0,0,0,0,0,0,0,0,0,16,0,0,0,0,0,\\ &0,0,0,0,0,0,0,0,0,0,0,0,0,0,0,0,0,0,0,0,176,0,0,0,0,0,0,0,0,0,0,0,0).\end{aligned}$$

Other pairs of sequences in $\mathcal{S}$ have the same cross-correlation property concerning the ZCZ.

### *B. Perfect Sequences and Orthogonal Sequence Sets*

The construction presented in III-A depends on the existence of perfect sequences and complete orthogonal sequence sets. In this subsection, we give a comment on their existence.

For an arbitrary integer $n \geq 3$, there is always a perfect sequence of length $n$ [29], and for some special $n$, there are many methods to construct them [30], [31]. However, in these constructions the number of phases of the sequences increases with the sequence length. For a fixed (and ideally, small) number of phases, we have few knowledge on the existence of perfect sequences of arbitrary length. For example, up to now only in length $2, 4, 8$ and $16$ have quadriphase perfect sequences been found.

For ternary (but not three-phase) sequences with entries in $\{0, \pm 1\}$, a lot of perfect sequences have been constructed [32], [33], [34], [35], [36]. More generally, in [37] an analytical method is proposed to find unimodular perfect sequence of any length.

For binary sequence set, the case is different. The following formula (22) is expected to hold for binary $(N, M; Zcz)$-ZCZ sets where $M \geq 2$ [11]:

$$Zcz \leq \frac{N}{2M}. \tag{22}$$

*Proposition 2:* Let $M \geq 2$. If Eq. $(22)$ holds for any binary $(N, M; Zcz)$-ZCZ set, then binary perfect sequences exist only in length 4.

The proof of this proposition is presented in Appendix IV.

Orthogonal sequence sets can be derived from unitary matrices and Hadamard matrices [20]. Hadamard matrices are a type of square $\{-1, 1\}$-matrices and exist for many orders $n$. Let $q = 0$ or $q$ be an odd prime, and $t$ and $l$ be arbitrary positive integers such that $n = 2^l(q^t+1) \equiv 0 \bmod 4$. Then there is always a Hadamard matrix of order $n$, and it can be constructed by Paley's method. A unitary matrix can be obtained by multiplying each entry of a Hadamard matrix of order $n$ by a same number $\frac{1}{\sqrt{n}}$. For an arbitrary positive integer $n$, a ternary complete orthogonal set of $n$ sequences can be constructed from Hadamard matrices [38].

Thus, although it can generate low-phase ZCZ sets, the method presented in the previous subsection is more suitable for constructing high-phase and ternary ZCZ sets.

### *C. Constructing Sets of ZCZ Sequence With Longer ZCZ and Larger Set Size*

Due to the sparsity of known low-phase perfect sequences, the low-phase ZCZ sets constructed in Section III-A do not possess a longer ZCZ and a larger family size simultaneously. In this subsection, we utilize the basic construction in Procedure 1 to present two methods, by combining which we can obtain ZCZ sets with both longer ZCZ and larger family size from known ZCZ sets.

Let $d$ be a fixed positive integer and $\mathcal{C}$ be a set of $l$ sequences $\mathbf{c}_k$ $(0 \le k < l)$, where $\mathbf{c}_k = (c_{k,0}, c_{k,1}, \cdots, c_{k,m-1})$. For each given $\mathbf{c}_k$, we define an ordered sequence set $\mathcal{A}^k = \{\mathbf{a}_0^k, \mathbf{a}_1^k, \cdots, \mathbf{a}_{n-1}^k\}$ where $\mathbf{a}_i^k = (a_{i,0}^k, \cdots, a_{i,m-1}^k)$ and

$$a_{i,j}^k = c_{k,(j(n+d)+i+d\lfloor\frac{i+1}{n}\rfloor) \bmod m}. \tag{23}$$

By Procedure 1, for any $0 \le k < l$, we use $\mathcal{A}^k$ and a fixed complete orthogonal set $\mathcal{B}$ of $n$ sequences to produce a sequence set $\mathcal{S}^k$. We combine the sets $\mathcal{S}^0, \mathcal{S}^1, \cdots, \mathcal{S}^{l-1}$ to get a larger set $\bigcup\limits_{0 \le k < l} \mathcal{S}^k$, which has $ln$ sequences.

*Theorem 3:* Let $\mathcal{C}$ be an $(m, l; Zcz)$-ZCZ set, $d$ be an integer with $0 \le d < Zcz$, and $m$ be relatively prime to $n + d$. Set

$$\begin{cases} r_0 = \lfloor \frac{Zcz-d}{n+d} \rfloor; \\ s_0 = \min\{n-1, Z_{CZ} - d - (n+d)r_0\}. \end{cases} \tag{24}$$

Then $\bigcup\limits_{0 \le k < l} \mathcal{S}^k$ is an $(mn, ln; r_0 n + s_0)$-ZCZ set. Furthermore,

(1) Assume $d = 0$ and $\gcd(m, n) = 1$. Then $\bigcup\limits_{0 \le k < l} \mathcal{S}^k$ is an $(mn, ln; Zcz)$-ZCZ set.

(2) Assume $l=1$ and $Zcz=m-1$, i.e., $\mathcal{C}$ consists of a single perfect sequence. If $n>Zcz$, then $\mathcal{S}^0$ is an $(mn,n;m-1-d)$-ZCZ set.

The proof of this theorem is presented in Appendix V.

The hypothesis on the relation between $m$ and $n$ in Theorem 3 is very weak, so it is convenient to apply the theorem to construct large ZCZ sets from the small ones. For the case that $m$ and $n$ are relatively prime, we can take $d=0$, thus getting an $(mn,n;m-1)$-ZCZ set which achieves the bound (7) by Theorem 3(2), provided a perfect sequence of length $m$ and a complete orthogonal set of $n$ sequences exist. For the case that $m$ and $n$ are not relatively prime, some ZCZ sequences are still constructed by our method, which extends Theorem 3 of [19]. For example, if $\mathcal{C}$ consists of a single perfect sequence of length $m$, and assume $n>m-1$ and $\gcd(m,n+1)=1$, then $\mathcal{S}^0$ is an $(mn,n;m-2)$-ZCZ set. It is an improvement on work of Torii *et al.* [20] since the divisibility condition $n|m$ or $m|n$ is not necessarily required. Some new quadriphase ZCZ sequence sets with $(L,M;Zcz)=(96,12;6),(320,20;14),(384,24;14)$ can be constructed by applying Theorem 3(2).

*Example 5:*

Suppose that $m=8,n=12,d=1$, a perfect quadiphase sequence $\mathbf{a}=(01022122)$ whose entries represent the same elements as those in Examples 3 and 4, and $\mathcal{B}$ is the complete orthogonal set in Example 4. By Theorem 3(2), we can obtain a quadiphase $(96,12;6)$-ZCZ set $\mathcal{S}$. We actually only show the sequences $\mathbf{s}_0$ and $\mathbf{s}_1$.

$$\begin{aligned}\mathbf{s}_0 = (&012221002120120010003203020122232200203022300003\\ &021020120010210021202300322210203022222322032020230).\\ \mathbf{s}_1 = (&030223000322102012001001002120230002221020302223\\ &223220320230012221002120120010003202020122232202032).\end{aligned}$$

For example, the absolute values of autocorrelation function of $\mathbf{s}_0$ are given by

$$\begin{aligned}|R_{\mathbf{s}_0}(\tau)| = (&96,0,0,0,0,0,0,8,16,0,0,0,0,0,16,8,0,0,0,0,0,0,16,8,0,0,0,0,0,24,\\ &16,0,0,0,0,0,0,8,0,0,0,0,0,0,16,8,0,0,0,0,0,8,16,0,0,0,0,0,0,8,0,0,0,0,\\ &0,0,16,24,0,0,0,0,0,8,16,0,0,0,0,0,0,8,16,0,0,0,0,0,16,8,0,0,0,0,0,0).\end{aligned}$$

The other sequences in $\mathcal{S}$ have the same autocorrelation property concerning the ZCZ. The absolute values of cross-correlation function of $\mathbf{s}_0$ and $\mathbf{s}_1$ are given by

$$
\begin{aligned}
|R_{\mathbf{s}_0,\mathbf{s}_1}(\tau)| = \; & 0,0,0,0,0,0,0,24,16,0,0,0,0,0,0,24,0,0,0,0,0,0,0,8,0,0,0,0,0, \\
& 8,32,0,0,0,0,0,0,24,0,0,0,0,0,0,16,24,0,0,0,0,0,8,0,0,0,0,0,0,0,88,0,0, \\
& 0,0,0,0,0,8,0,0,0,0,0,8,0,0,0,0,0,0,0,8,0,0,0,0,0,0,32,24,0,0,0,0,0,0).
\end{aligned}
$$

Other pairs of sequences in $\mathcal{S}$ have the same cross-correlation property concerning the ZCZ.

*Example 6:*

Let $\mathcal{C}$ be the quadriphase $(64, 4; 14)$-ZCZ set shown in Example 3(1). Take $d = 1$ and $n = 16$. Let $H$ be the matrix in Eq. (15) and $\mathcal{B}$ consist of row sequences of $H \otimes H \otimes H \otimes H$ which is a Hadamard matrix of order 16, where $\otimes$ denotes the Kronecker product. Applying Theorem 3, we obtain a quadriphase $(1024, 64; 13)$-ZCZ set.

Theorem 3 provides a method to construct a ZCZ set with larger family size from that with smaller family size. The following theorem is directly derived from the basic construction in Procedure 1. It is a method to construct a ZCZ set with longer ZCZ from a set with shorter ZCZ.

*Theorem 4:* If $A$ is an $(m, n; Zcz)$-ZCZ set, then $\mathcal{S}$ constructed by Procedure 1 is an $(mn, n; nZcz)$-ZCZ set.

The proof of Theorem 4 is presented in Appendix VI.

A variant of Theorem 4 is proved in [20], where the divisibility condition $n|m$ or $m|n$ is assumed and the variant is used to iteratively construct new ZCZ set from the old ones. Our variant of Theorem 4 removes this divisibility assumption, and hence it is applicable to construct more ZCZ sets.

The application of Theorems 3 and 4 depends not on the existence of perfect sequences but on that of complete orthogonal sequence sets. This relaxes restriction on the application of the two theorems since complete orthogonal sequence sets are easily constructed. Applying them to known ZCZ sets, we will obtain new sets with both longer ZCZ and larger set size. Moreover, for the construction of some even-phase, say $p$-phase, ZCZ sets, we choose some specific $n$ such that $n = 2^l(q^t + 1) \equiv 0 \bmod 4$ as mentioned in III-B; since there always exists a Hadamard

TABLE I

SOME KNOWN QUADRIPHASE (N,M,ZCZ)-ZCZ SEQUENCES SETS FOR $M \leq \frac{N}{M} \leq 16$

| $\frac{N}{M}$ | 4 | 4 | 8 | 8 | 8 | 16 | 16 | 16 | 16 | 16 |
|---|---|---|---|---|---|---|---|---|---|---|
| $M$ | 2 | 4 | 2 | 4 | 8 | 2 | 4 | 8 | 12 | 16 |
| $Zcz$ | 2 | 2 | 6 | 6 | 6 | 14 | 14 | 14 | $12^{\triangle}$ | 14 |

TABLE II

SOME KNOWN QUADRIPHASE (N,M,ZCZ)-ZCZ SEQUENCES SETS FOR $8 \leq \frac{N}{M} \leq 16 < M \leq 36$

| $\frac{N}{M}$ | 8 | 16 | 8 | 16 | 8 | 16 | 8 | 16 | 8 | 16 |
|---|---|---|---|---|---|---|---|---|---|---|
| $M$ | 20 | 20 | 24 | 24 | 28 | 28 | 32 | 32 | 36 | 36 |
| $Zcz$ | $6^{\triangle}$ | $14^{\triangle}$ | 6 | $14^{\triangle}$ | $6^{\triangle}$ | $14^{\triangle}$ | 6 | 14 | $6^{\triangle}$ | $14^{\triangle}$ |

matrix of order $n$, the new set constructed for this kind of parameter $n$ has the same number of phases as the old one, that is, it is still of $\frac{2p}{\gcd(p,2)} = p$-phase.

*Example 7:* Suppose that $\mathcal{A}$ is the quadriphase $(1024, 64; 13)$-ZCZ set constructed in Example 6. By Theorem 4 we get a quadriphase $(65536, 64; 832)$-ZCZ set since there exists a complete orthogonal set of 64 sequences derived from a Hadamard matrix of order 64.

### *D. The comparison of parameters on known constructions*

In this correspondence, some new ZCZ sequences obtained can reach the bound (7), as shown in Theorems 1, 2 and 3. But there are very strict limitations on the parameters: the length of perfect sequences and the length of the shift sequences and the order of the Hadamard matrices.

Those quadriphase $(N, M; Zcz)$-ZCZ sequences obtained in this correspondence can not reach the bound (7), however, for given $N$ and $M$, their zero correlation zone $Zcz$ are the maximal among all known results about quadriphase $(N, M; Zcz)$-ZCZ sequences. Tables I and II summarize some of the best known quadriphase $(N, M; Zcz)$-ZCZ sequences, and the symbol "$\triangle$" mean that the quadriphase $(N, M; Zcz^{\triangle})$-ZCZ sequences can be constructed by our method but not by others. Utilizing Theorems 3 and 4 to these sequences, some new sequence sets are obtained.

## IV. Sequences With Low Cross Correlation

Some sequence sets with low cross correlation are found, e.g., those from Gold-pair construction and interleaved construction [39], [17], [18], Kasami sequences [40], and bent function sequences [41]. Using the basic method in Section II-B, we can also construct sequences with low out-of-phase autocorrelation and cross-correlation.

Let $\mathcal{A} = \{L^{e_0}(\mathbf{a}), L^{e_1}(\mathbf{a}), \cdots, L^{e_{n-1}}(\mathbf{a})\}$ be an ordered set of shift equivalent sequences, and $m, n, \mathcal{B}$ and $\mathcal{S}$ be as in Procedure 1. Set

$$N_{r,s} = |\{0 \leq j < n |\ e_{j+s} - e_j + r \equiv 0 \bmod m\}|$$

and

$$N_0 = \max\{N_{r,s}|\ 0 \leq r < m,\ 0 \leq s < n\}.$$

*Theorem 5:* Assume $\mathbf{a}$ is a perfect sequence and the absolute value of each entry in the sequences of set $\mathcal{B}$ does not exceed 1. Then $\mathcal{S}$ is an $[mn, n; N_0 E_{\mathbf{a}}]$ sequence set. Furthermore,

(1) If $m \geq n$ and the shift sequence $\mathbf{e} = (e_0, e_1, \cdots, e_{n-1})$, where $0 \leq e_i < m$ for all $i$, satisfies

$$|\{(e_{j+s} - e_j) \bmod m \mid 0 \leq j < n - s\}| = n - s \qquad (25)$$

for any $1 \leq s < n$, then $\mathcal{S}$ is an $[mn, n; 2E_a]$ sequence set.

(2) If $m \geq 2n$ and $\mathbf{e} = (e_0, e_1, \cdots, e_{n-1})$, where $0 \leq e_i < n$ for all $i$, satisfies

$$|\{(e_{j+s} - e_j) \bmod n \mid 0 \leq j < n - s\}| = n - s \qquad (26)$$

for any $1 \leq s < n$, then $\mathcal{S}$ is an $[mn, n; 2E_a]$ sequence set.

The proof of this theorem is presented in Appendix VII.

Gong provides two methods in [18] to construct shift sequences $\mathbf{e} = (e_0, e_1, \cdots, e_{n-1})$ satisfying equality $(25)$. In her methods, $0 \leq e_i < n$ for $0 \leq i < n$ and the length of the perfect sequence is a prime or is of the form $p^l - 1$ for some prime $p$. If we take $m = n$ in Theorem 5(1), then new sequence set can be constructed by Gong's methods. However, this reduces applicability of Procedure 1 since perfect sequences with such a length may exist in small quantities.

*Example 8:* Let $m = n = 8$, $\mathbf{e} = (0, 5, 6, 5, 7, 7, 3, 6)$, $\mathbf{a} = (00120210)$ be a quadriphase perfect sequence, and $\mathcal{B}$ be a complete orthogonal set derived from $H \otimes H \otimes H$. Then $\mathcal{S}$ constructed by Procedure 1 is a quadriphase $[64, 8; 16]$ sequence set. We list the first two sequences in $S$ as follows:

$$\mathbf{s}_0 = (0212002101010000100000202000111000102201212100021202220000201112)$$

$$\mathbf{s}_1 = (0010022303030202120202222202131202122003232302001000200202221310)$$

The absolute values of autocorrelation function of $\mathbf{s}_0$ and the absolute values of cross-correlation function of $\mathbf{s}_0$ and $\mathbf{s}_1$ are given by

$$| R_{\mathbf{s}_0}(\tau) |= (64, 8, 16, 0, 8, 16, 8, 8, 0, 8, 8, 8, 16, 16, 8, 8, 0, 0, 16, 8, 0, 8, 0, 16, 0, 8, 0, 8, 8,$$
$$0, 8, 8, 0, 8, 8, 0, 8, 8, 0, 8, 0, 16, 0, 8, 0, 8, 16, 0, 0, 8, 8, 16, 16, 8, 8, 8, 0, 8, 8, 16, 8, 0, 16, 8)$$

and

$$| R_{\mathbf{s}_0,\mathbf{s}_1}(\tau) |= (0, 8, 16, 0, 8, 0, 8, 8, 0, 8, 8, 8, 0, 16, 8, 8, 0, 0, 16, 8, 0, 8, 0, 0, 0, 8, 0, 8, 8,$$
$$0, 8, 8, 0, 8, 8, 0, 8, 8, 0, 8, 0, 0, 0, 8, 0, 8, 16, 0, 0, 8, 8, 16, 0, 8, 8, 8, 0, 8, 8, 0, 8, 0, 16, 8, ).$$

Similarly, let $m = 16$, $n = 8$, $\mathbf{a}$ be the quadriphase perfect sequence of length 16 listed in Example 3, and $\mathbf{e}$ and $\mathcal{B}$ be the same as above, then by Theorem 5(2) we can obtain a quadriphase $[128, 8; 32]$ sequence set.

Theorem 5(2) is especially suitable for constructing ternary sequence sets in the sense that there are a lot of ternary perfect sequences [36] and a ternary complete orthogonal set of $n$ sequences exists for an arbitrary positive integer $n$ [38].

Let $\mathbf{a} = (a_0, a_1, \cdots, a_{m-1})$ be a ternary perfect sequence, and $\mathcal{B}$ be a ternary complete orthogonal set of $n$ sequences, where $n \in [\frac{m}{4}, \frac{m}{2}]$ is a prime or is of the form $p^s - 1$ where $p$ is a prime. It is easy to find such $n$ and $\mathcal{B}$ for a given $m \geq 4$. We can use Gong's methods in [18] to find a shift sequence $\mathbf{e} = (e_0, e_1, \cdots, e_{n-1})$ satisfying Eq. $(26)$, and an $[mn, n; 2E_{\mathbf{a}}]$ sequence set $\mathcal{S}$ is obtained by Theorem 5(2). Different from Gong's $[n^2, n; 2n + 3]$ set [18], our construction provides sequence sets with different choices of parameters of set size and sequence length.

*Example 9:* Let $m = 13$, $n = 8$, $\mathbf{e} = (0, 5, 6, 5, 7, 7, 3, 6)$,

$$\mathbf{a} = (-1, -1, -1, -1, 0, 1, -1, 1, 0, 0, -1, 0, 1)$$

be a ternary perfect sequence, and $\mathcal{B}$ be the complete orthogonal set in Example 8. By Theorem 5 we get a ternary $[104, 8; 18]$ sequence set $\mathcal{S}$. The first two sequences in $\mathcal{S}$ are shown as

$$\begin{aligned}\mathbf{s}_0 = (&- + - + + + - - - - + - 000 + - + 0 + 00 + 0 - 000 - - - 000- \\ &000 + - + -0 - + + 00 - 0 + 0 - -0 + + + - + - - - - 0 - - - -- \\ &0 - 0 - - - - - + - - - - - 00 - -0 - 0 - + + -0 + 0 + 0 - - - +)\end{aligned}$$

and

$$\begin{aligned}\mathbf{s}_1 = (&- - - - + - - + - + + + 000 - - - 0 - 00 + 0 - 000 - + - 000- \\ &000 + + + +0 + + - 00 - 0 + 0 - +0 - + - - - - + - + 0 + - + -+ \\ &0 + 0 + - + - + + + - + - + 00 - +0 + 0 + + - -0 + 0 + 0 - + - -)\end{aligned}$$

where $+$ and $-$ represent $+1$ and $-1$, respectively. The absolute values of auto- and cross-correlation of $\mathbf{s}_0$ and $\mathbf{s}_1$ are given by

$$\begin{aligned}|R_{\mathbf{s}_0}(\tau)| = |R_{\mathbf{s}_1}(\tau)| = (&72, 9, 18, 0, 9, 18, 9, 9, 0, 9, 9, 9, 9, 9, 9, 9, 0, 0, 9, 9, 0, 9, 0, 9, 0, 0, 0, \\ &0, 9, 0, 0, 0, 0, 9, 9, 0, 0, 9, 0, 9, 0, 9, 0, 0, 0, 0, 9, 0, 0, 0, 0, 9, 18, 9, 0, 0, 0, 0, 9, 0, 0, 0, 0, 9, 0, 9, \\ &0, 9, 0, 0, 9, 9, 0, 0, 0, 0, 9, 0, 0, 0, 0, 9, 0, 9, 0, 9, 9, 0, 0, 9, 9, 9, 9, 9, 9, 9, 0, 9, 9, 18, 9, 0, 18, 9)\end{aligned}$$

and

$$\begin{aligned}|R_{\mathbf{s}_0,\mathbf{s}_1}(\tau)| = |R_{\mathbf{s}_1,\mathbf{s}_0}(\tau)| = (&0, 9, 18, 0, 9, 0, 9, 9, 0, 9, 9, 9, 9, 9, 9, 9, 0, 0, 9, 9, 0, 9, 0, 9, 0, 0, 0, \\ &0, 9, 0, 0, 0, 0, 9, 9, 0, 0, 9, 0, 9, 0, 9, 0, 0, 0, 0, 9, 0, 0, 0, 0, 9, 18, 9, 0, 0, 0, 0, 9, 0, 0, 0, 0, 9, 0, 9, \\ &0, 9, 0, 0, 9, 9, 0, 0, 0, 0, 9, 0, 0, 0, 0, 9, 0, 9, 0, 9, 9, 0, 0, 9, 9, 9, 9, 9, 9, 9, 0, 9, 9, 0, 9, 0, 18, 9)\end{aligned}$$

## V. CONCLUSION

In this correspondence we consider how to construct sequence sets with zero-correlation zone or with low correlation. We present a general method to generate new sequence sets with these correlation properties. We prove two theorems (Theorems 3 and 4), and by applying them to known ZCZ sequence sets, we can obtain new ZCZ sets with both longer ZCZ and larger family size. More sets of sequences seem to be likely to be generated if appropriate parameters are carefully chosen in Procedure 1.

## APPENDIX I
## PROOF OF PROPOSITION 1

*Proof:* Let $\mathbf{u}$ be the sequence associated with the ordered set $\mathcal{A}$ (see Procedure 1 (2)), and $T = (T_0, T_1, \cdots, T_{n-1})$ the matrix form of $L^\tau(\mathbf{u})$. By Procedure 1 (2), for $0 \le i < n$, the $i$-th entry in the sequence $L^\tau(\mathbf{u})$ is $a_{(s+i)\bmod n,(r+\varphi(s+i))\bmod m}$ and it is exactly the first element of $T_i$. So,

$$T_i = L^{(r+\varphi(s+i))\bmod m}(\mathbf{a}_{s+i-\varphi(s+i)n}). \tag{27}$$

Since the $i$-th column of the matrix form of $\mathbf{s}_k$ is $b_{k,i}\mathbf{a}_i$, the $i$-th column of the matrix form of $L^\tau(\mathbf{s}_k)$ is

$$L^{(r+\varphi(s+i))\bmod m}(b_{k,s+i-\varphi(s+i)n}\mathbf{a}_{s+i-\varphi(s+i)n}). \tag{28}$$

Thus, one has

$$\begin{aligned} & R_{\mathbf{s}_h,\mathbf{s}_k}(\tau) \\ = & \sum_{i=0}^{n-1}\sum_{l=0}^{m-1}(a_{i,l}b_{h,i})(a^*_{s+i-\varphi(s+i)n,(l+r+\varphi(s+i))\bmod m}b^*_{k,s+i-\varphi(s+i)n}) \\ = & \sum_{i=0}^{n-1}\sum_{l=0}^{m-1}(a_{i,l}a^*_{s+i-\varphi(s+i)n,(l+r+\varphi(s+i))\bmod m})(b_{h,i}b^*_{k,s+i-\varphi(s+i)n}) \\ = & \sum_{i=0}^{n-1} d_i R_{\mathbf{a}_i,\mathbf{a}_{s+i-\varphi(s+i)n}}(r+\varphi(s+i)) \end{aligned} \tag{29}$$

where $d_i = b_{h,i}b^*_{k,s+i-\varphi(s+i)n}$.

## Appendix II
## Proof of Theorem 1

*Proof:* Set $t_i = e_{i+s} - e_i + r$. Then by Eq. (18),

$$R_{\mathbf{s}_h,\mathbf{s}_k}(\tau) = \sum_{i=0}^{n-1} d_i R_{\mathbf{a}}(t_i).$$

where $\tau = rn + s$.

(i) Assume $0 \le r \le r_0$ and $s = 0$. Then

$$R_{\mathbf{s}_h,\mathbf{s}_k}(rn) = R_{\mathbf{a}}(r)\sum_{i=0}^{n-1} d_i = R_{\mathbf{a}}(r)R_{\mathbf{b}_h,\mathbf{b}_k}(0).$$

Thus $R_{\mathbf{s}_h,\mathbf{s}_k}(0) = E_{\mathbf{a}} \cdot n$ for $h = k$ and $r = 0$, and $R_{\mathbf{s}_h,\mathbf{s}_k}(rn) = 0$ otherwise.

(ii) Assume $0 \le r \le r_0$ and $s > 0$. If $i + s < n$,

$$0 \le r < e_{i+s} - e_i + r \le e_{n-1} - e_0 + r_0 \le m;$$

If $i+s \geq n$,

$$\begin{aligned} -m &< e_{i+s} - e_i + r \\ &= 1 + e_{i+s-n} - e_i + r \\ &\leq 1 + e_{i-1} - e_i + r \\ &\leq 1 + e_{i-1} - e_i + r_0 \\ &\leq 0. \end{aligned}$$

So, either $0 < t_i \leq m$ or $-m < t_i \leq 0$ holds, and if the equality holds then $s = n-1$ and $r = r_0$. Thus, for any $0 < \tau \leq r_0 n + n - 2$, either $0 < t_i < m$ or $-m < t_i < 0$ holds for all $i \in \{0, 1, \cdots, n-1\}$, and consequently, $R_{\mathbf{s}_h, \mathbf{s}_k}(\tau) = 0$.

Combining the above two cases, we have $R_{\mathbf{s}_h, \mathbf{s}_k}(\tau) = 0$ for $0 < \tau \leq r_0 n + n - 2$ and for $\tau = 0$ and $h \neq k$. By

$$R_{\mathbf{s}_h, \mathbf{s}_k}(-\tau) = R_{\mathbf{s}_k, \mathbf{s}_h}(\tau)^*, \tag{30}$$

we conclude that $\mathcal{S}$ is an $(mn, n; r_0 n + n - 2)$-ZCZ set.

(1) Assume $n \mid (m+1)$. Equalities

$$e_{i+1} - e_i = \frac{m+1}{n} \text{ for } 0 \leq i \leq n-2$$

imply $r_0 = \frac{m+1}{n} - 1$, and $\mathcal{S}$ is an $(mn, n; m-1)$-ZCZ set.

(2) Assume $n \mid m$ and

$$\frac{(i_1 - i_2)m}{n} \leq e_{i_1} - e_{i_2} \leq \frac{(i_1 - i_2)m}{n} + 1$$

for $0 \leq i_2 < i_1 \leq n-1$. Then

$$\frac{m}{n} - 1 \leq m + e_0 - e_{n-1} \leq \frac{m}{n}$$

and

$$\frac{m}{n} - 1 \leq e_i - e_{i-1} - 1 \leq \frac{m}{n}$$

for $1 \leq i \leq n-1$.

If $m + e_0 - e_{n-1} = \frac{m}{n}$ and

$$e_i - e_{i-1} - 1 = \frac{m}{n} \text{ for } 1 \leq i \leq n-1,$$

then

$$e_{n-1} - e_0 = \sum_{j=1}^{n-1}(e_j - e_{j-1}) = \frac{(n-1)m}{n} + (n-1)$$

and it conflicts to $m+e_0-e_{n-1}=\frac{m}{n}$. So either

$$m+e_0-e_{n-1}=\frac{m}{n}-1 \quad \text{or} \quad e_{i_0}-e_{i_0-1}-1=\frac{m}{n}-1$$

holds for some $i_0$. Thus $r_0=\frac{m}{n}-1$ and $\mathcal{S}$ is an $(mn,n;m-2)$-ZCZ set.

If $\mathbf{e}=\mathbf{e}^{(i)}$, it is easy to check that $r_0=\frac{m}{n}-1$ and $\mathcal{S}$ is an $(mn,n;m-2)$-ZCZ set.

## APPENDIX III
## PROOF OF THEOREM 2

*Proof:* It is easy to check that $\mathcal{S}$ is an $(mn,n;0)$-ZCZ set if $m=n$, so we only need to consider the case of $n<m$.

By Eq. (18), for $r=0$ and $\tau=s$,

$$R_{\mathbf{s}_h,\mathbf{s}_k}(\tau)=\sum_{i=0}^{n-1}d_iR_{\mathbf{a}}(e_{i+s}-e_i).$$

By $e_i=m-1-i$, we have

$$e_{i+s}-e_i\equiv\begin{cases}-s & \text{if } i+s<n;\\ n+1-s & \text{if } i+s\geq n.\end{cases} \tag{31}$$

Thus, for $0<s<n$,

$$R_{\mathbf{a}}(e_{i+s}-e_i)=0 \text{ for all } 0\leq i\leq n-1;$$

for $r=1$ and $s=0$,

$$R_{\mathbf{s}_h,\mathbf{s}_k}(\tau)=\sum_{i=0}^{n-1}d_iR_{\mathbf{a}}(1)=0.$$

For $h\neq k$,

$$R_{\mathbf{s}_h,\mathbf{s}_k}(0)=\sum_{i=0}^{n-1}d_iR_{\mathbf{a}}(0)=R_{\mathbf{a}}(0)R_{\mathbf{b}_h,\mathbf{b}_k}(0)=0.$$

By Eq. (30), $\mathcal{S}$ is an $(mn,n,n \bmod m)$-ZCZ set.

## APPENDIX IV
## PROOF OF PROPOSITION 2

*Proof:* Let $\mathbf{a} = (a_0, a_1, \cdots, a_{m-1})$ be a binary perfect sequence, $m \geq 4$. Take a positive integer $n$ such that $n = 2^t \leq m$ for some integer $t \geq 1$, and a complete orthogonal set $\mathcal{B}$ derived from a Hadamard matrix of order $2^t$.

(i) If $n = m$, we use $\mathbf{a}$ and $\mathcal{B}$ to construct an $(m^2, m; m-2)$-ZCZ set [20]. Then by the bound (22), $m - 2 \leq \frac{m}{2}$, i.e., $m \leq 4$.

(ii) If $n < m$, we assume $\frac{m}{2^{s+1}} \leq n < \frac{m}{2^s}$ for some $s \geq 0$.

(ii.1) If $n = \frac{m}{2^{s+1}}$, then by Theorem 1, an $(mn, n; m-2)$-ZCZ set is constructed. Similarly as in (i), one has $m \leq 4$.

(ii.2) If $\frac{m}{2^{s+1}} < n < \frac{m}{2^s}$, then a $(2^s mn, 2^s n; 2^s n)$-ZCZ set is constructed form $\mathbf{a}$ and $\mathcal{B}$ by Theorem 2. In such a case, $\frac{m}{2} < 2^s n < m$, and it contradicts the assumed bound (22). Thus $m \leq 4$.

## APPENDIX V
## PROOF OF THEOREM 3

*Proof:* We only need to prove that

$$R_{\mathbf{s}_i^k, \mathbf{s}_{i'}^{k'}}(\tau) = 0 \text{ for } 1 \leq \tau \leq r_0 n + s_0$$

and

$$R_{\mathbf{s}_i^k, \mathbf{s}_{i'}^{k'}}(0) = 0 \text{ for } k \neq k' \text{ or } i \neq i'.$$

It is easy to check that if $\mathbf{s}_i^k$ and $\mathbf{s}_{i'}^{k'}$ are the same sequences, then $k = k'$ and $i = i'$. So, $\bigcup_{0 \leq k < l} \mathcal{S}^k$ has $ln$ different sequences. By $(9)$ and $(14)$, one has

$$s_{i,t}^k = a^k_{t \bmod n, \lfloor \frac{t}{n} \rfloor} b_{i, t \bmod n} = c_{k, (t + d \lfloor \frac{t+1}{n} \rfloor) \bmod m} b_{i, t \bmod n}$$

and

$$\begin{aligned} & R_{\mathbf{s}_i^k,\mathbf{s}_{i'}^{k'}}(\tau) \\ =& \sum_{t=t_1n+t_2=0}^{mn-1} s_{i,t}^k s_{i',(t+\tau)\bmod mn}^{k'*} \\ =& \sum_{t_2=0}^{n-1}\sum_{t_1=0}^{m-1}(b_{i,t_2}b^*_{i',(s+t_2)\bmod n})(a^k_{t_2,t_1}a^{k'*}_{(s+t_2)\bmod n,(t_1+r+\varphi(s+t_2))\bmod m}) \\ =& \sum_{t_2=0}^{n-1}\sum_{t_1=0}^{m-1}(b_{i,t_2}b^*_{i',(s+t_2)\bmod n})(c_{k,p}c^*_{k',q}) \\ =& \sum_{t_2=0}^{n-1} d_{t_2}R_{\mathbf{c}_k,\mathbf{c}_{k'}}(\tau+d(r+\lfloor\tfrac{s+t_2+1}{n}\rfloor-\lfloor\tfrac{t_2+1}{n}\rfloor)) \end{aligned} \tag{32}$$

where $\tau = rn + s$, $d_{t_2} = b_{i,t_2}b^*_{i',s+t_2}$, and the subscripts

$$p = (t_1(n+d) + i_2 + d\lfloor\frac{t_2+1}{n}\rfloor)\bmod m$$

and

$$q = ((t_1 + r + \varphi(s+t_2))(n+d) + (s+t_2)\bmod\, n + d\lfloor\frac{((s+t_2)\bmod n)+1}{n}\rfloor)\bmod\, m.$$

The last equality in Eq. (32) holds since

$$\begin{aligned} q-p &= (t_1+r+\varphi(s+t_2))(n+d) + s + t_2 - \varphi(s+t_2)n \\ &\quad + d\lfloor\tfrac{s+t_2-\varphi(s+t_2)n+1}{n}\rfloor - (t_1(n+d)+t_2+d\lfloor\tfrac{t_2+1}{n}\rfloor) \\ &= (t_1+r)(n+d) + s + t_2 + d\lfloor\tfrac{s+t_2+1}{n}\rfloor - (t_1(n+d)+t_2+d\lfloor\tfrac{t_2+1}{n}\rfloor) \\ &= \tau + dr + d\lfloor\tfrac{s+t_2+1}{n}\rfloor - d\lfloor\tfrac{t_2+1}{n}\rfloor. \end{aligned}$$

and $\gcd(m, n+d) = 1$ implies that, for any fixed $0 \le t_2 < n$,

$$|\{(t_1(n+d) + t_2 + d\lfloor\frac{t_2+1}{n}\rfloor)\mathrm{modm} \mid 0 \le t_1 < m\}| = m.$$

By (18),

$$R_{\mathbf{s}_i^k,\mathbf{s}_{i'}^{k'}}(0) = R_{\mathbf{c}^k,\mathbf{c}^{k'}}(0)R_{\mathbf{b}_i,\mathbf{b}_{i'}}(0) = 0$$

if $k \ne k'$ or $i \ne i'$. When $1 \le \tau \le r_0n + s_0$,

$$\begin{aligned} 1 &\le \tau + d(r + \lfloor\tfrac{s+t_2+1}{n}\rfloor - \lfloor\tfrac{t_2+1}{n}\rfloor) \\ &\le r_0n + s_0 + dr_0 + d \\ &\le Zcz. \end{aligned}$$

Thus, $R_{\mathbf{s}_i^k,\mathbf{s}_{i'}^{k'}}(\tau) = 0$ for $1 \le \tau \le r_0n + s_0$.

## APPENDIX VI
## PROOF OF THEOREM 4

*Proof:* Since $\mathcal{B}$ is a complete orthogonal sequence set, any two sequences in $\mathcal{S}$ are different. Let $\tau = rn + s$ and $\mathrm{s}_h, \mathrm{s}_k \in \mathcal{S}$.

(i) For $0 < r < Zcz$ and $0 \leq s < n$, since $\mathcal{A}$ is an $(m, n; Zcz)$-ZCZ set,

$$R_{\mathbf{a}_i, \mathbf{a}_{s+i-\varphi(s+i)n}}(r + \varphi(s+i)) = 0.$$

So by Proposition 1, $R_{\mathbf{s}_h, \mathbf{s}_k}(\tau) = 0$;

(ii) For $r = Zcz$ and $s = 0$, $\varphi(s+i) = 0$, and by Proposition 1, one also has $R_{\mathbf{s}_h, \mathbf{s}_k}(\tau) = 0$;

(iii) For $r = 0$ and any $0 < s < n$,

$$R_{\mathbf{a}_i, \mathbf{a}_{s+i-\varphi(s+i)n}}(\varphi(s+i)) = 0.$$

Thus,

$$R_{\mathbf{s}_h, \mathbf{s}_k}(\tau) = \sum_{i=0}^{n-1} d_i R_{\mathbf{a}_i, \mathbf{a}_{s+i-\varphi(s+i)n}}(\varphi(s+i)) = 0.$$

If $s = 0$ and $h \neq k$,

$$R_{\mathbf{s}_h, \mathbf{s}_k}(0) = \sum_{i=0}^{n-1} d_i R_{\mathbf{a}_i}(0) = E_{\mathbf{a}_0} \cdot R_{\mathbf{b}_h, \mathbf{b}_k}(0) = 0.$$

By Eq. (30), we conclude $\mathcal{S}$ is an $(mn, n; nZcz)$-ZCZ set.

## APPENDIX VII
## PROOF OF THEOREM 5

*Proof:* Obviously, $\mathcal{S}$ has $n$ different sequences.

$$|R_{\mathbf{s}_h, \mathbf{s}_k}(\tau)| = |\sum_{i=0}^{n-1} d_i R_{\mathbf{a}}(t_i)| \leq \sum_{i=0}^{n-1} |R_{\mathbf{a}}(t_i)| \leq N_0 E_{\mathbf{a}}.$$

where $t_i = e_{i+s} - e_i + r$.

(1) If

$$|\{(e_{i+s} - e_i) \text{mod } m \mid 0 \leq i < n - s\}| = n - s$$

hods for all $1 \leq s < n$, then

$$(e_{i_1+s} - e_{i_1}) \text{mod } m \neq (e_{i_2+s} - e_{i_2}) \text{mod } m$$

and

$$(e_{i_1+s} - e_{i_1} + r) \bmod m \neq (e_{i_2+s} - e_{i_2} + r) \bmod m$$

for $0 \leq i_1 \neq i_2 < n - s$. Thus, there is at most one integer $i_0$ $(0 \leq i_0 < n - s)$ such that $t_{i_0} \equiv 0 \bmod m$. Similarly, there is at most one integer $i'_0$ $(n - s \leq i'_0 < n)$ such that $t_{i'_0} \equiv 0 \bmod m$. Then, one has $N_0 \leq 2$.

(2) If

$$|\{(e_{i+s} - e_i) \bmod n \mid 0 \leq i < n - s\}| = n - s$$

for all $1 \leq s < n$, then

$$e_{i_1+s} - e_{i_1} \neq e_{i_2+s} - e_{i_2}$$

and

$$-m \leq -2n < t_{i_1} - t_{i_2} < 2n \leq m$$

for $0 \leq i_1 \neq i_2 < n - s$. Thus

$$(e_{i_1+s} - e_{i_1} + r) \bmod m \neq (e_{i_2+s} - e_{i_2} + r) \bmod m$$

and there is at most one integer $i_0$ $(0 \leq i_0 < n - s)$ such that $t_{i_0} \equiv 0 \bmod m$. Similarly, there is at most one integer $i'_0$ $(n - s \leq i'_0 < n)$ such that $t_{i'_0} \equiv 0 \bmod m$. Then, one has $N_0 \leq 2$.